\documentclass[aps,prl,reprint,twocolumn,superscriptaddress]{revtex4-2}
\usepackage{amsmath,amssymb}
\usepackage{graphicx}
\usepackage{overpic}
\usepackage{xcolor}
\usepackage{hyperref}
\graphicspath{{../pictures/}}

\begin{document}

\title{Lee-Yang Theory Guided Force Field Refinement Based on Phase Diagrams}

\author{Bin Jin}
\affiliation{School of Materials Science and Engineering, Peking University, Beijing 100871, China}

\author{Qijun Ye}
\affiliation{Interdisciplinary Institute of Light-Element Quantum Materials, Research Center for Light-Element Advanced Materials, and Collaborative Innovation Center of Quantum Matter, Peking University, Beijing 100871, China}
\affiliation{State Key Laboratory for Artificial Microstructure and Mesoscopic Physics, Frontier Science Center for Nano-optoelectronics and School of Physics, Peking University, Beijing 100871, China}
\affiliation{Peking University Yangtze Delta Institute of Optoelectronics, Nantong, Jiangsu 226010, China}

\author{Xinzheng Li}
\affiliation{Interdisciplinary Institute of Light-Element Quantum Materials, Research Center for Light-Element Advanced Materials, and Collaborative Innovation Center of Quantum Matter, Peking University, Beijing 100871, China}
\affiliation{State Key Laboratory for Artificial Microstructure and Mesoscopic Physics, Frontier Science Center for Nano-optoelectronics and School of Physics, Peking University, Beijing 100871, China}
\affiliation{Peking University Yangtze Delta Institute of Optoelectronics, Nantong, Jiangsu 226010, China}

\author{Kuang Yu}
\email{kuangyu.2025@bytedance.com}
\affiliation{ByteDance Seed --- AI for Science, Shenzhen Bay Technology Innovation Center, Shenzhen 518000, Guangdong, China}

\author{Shenzhen Xu}
\email{xushenzhen@pku.edu.cn}
\affiliation{School of Materials Science and Engineering, Peking University, Beijing 100871, China}

\begin{abstract}
We propose a general framework for automatic force field refinement guided by phase diagrams, grounded in Lee-Yang phase transition theory. The central idea is to directly use the partition function modulus as a phase-diagram-guided optimization target. Evaluating the modulus at points close to the real axis, where the Lee-Yang zeros are mostly associated with the phase transition, is more physically meaningful and avoids the numerical difficulty of explicitly solving for the zeros. This approach requires no system-specific order parameters or response properties for characterizing phase transition points, making it universal across various discontinuous phase transitions and material systems. We validate the method on refining parameters of a Lennard-Jones potential covering both gas-liquid and solid-liquid transitions, and a Cu embedded-atom method potential based on experimental melting curves. The refined force fields reproduce the target phase diagrams with significant improvement across all systems. For Cu, the refinement simultaneously improves predictions of enthalpy and heat capacity, which are observables beyond the optimization target. These results establish Lee-Yang theory as a practical tool for contemporary force field development.
\end{abstract}

\maketitle

Phase transitions are among the most fundamental phenomena in condensed matter physics and materials science~\cite{anderson1984basic,chaikin1995principles,stanley1987phase}, governing processes from the freezing of water to the melting of structural alloys. Molecular dynamics (MD) and Monte Carlo (MC) simulations have become indispensable tools for studying phase behavior at the microscopic level~\cite{frenkel2002understanding,tuckerman2010statistical}, yet their predictive power hinges critically on the quality of the interatomic potential, or force field, that describes the interactions between particles.

Recent advances in automatic differentiation have opened new possibilities for force field development. Two broad strategies exist for phase-diagram-guided force field optimization. One approach determines the phase transition point by locating the thermodynamic condition where the free energies of the two phases become equal, computing the free energy of each phase separately~\cite{fuchs2025refining,thaler2025chemtrain}. The alternative strategy, which we adopt here and in our prior work, is to directly simulate phase coexistence. The advantage of the coexistence approach is that it determines the self-consistent coexistence state directly from simulation.

Our prior work realized this coexistence-based strategy using the Differentiable Molecular Force Field (DMFF) framework~\cite{wang2023dmff}, demonstrating that force field parameters can be directly optimized against gas-liquid coexistence data~\cite{jin2026automatic}. This strategy employed density distribution as the optimization target, enabling successful refinement of Lennard-Jones~\cite{jones1924part1,jones1924part2,jones1931cohesion} and CO$_2$ TraPPE~\cite{potoff2001trappe} force fields by matching simulated and experimental coexistence curves. Extending this paradigm to solid-liquid transitions, however, reveals fundamental limitations.

The gas-liquid refinement strategy relied on density as a natural order parameter. In solid-liquid transitions, by contrast, density differences between phases are far smaller, rendering density-based optimization targets unreliable. More fundamentally, traditional methods require experimental information that extends beyond the $p$-$T$ phase diagram itself. Order parameters might be chosen to distinguish competing phases, introducing system-dependent choices that are not directly constrained by experiment. Approaches based on thermodynamic properties use derivatives of the partition function, including first-order quantities such as energy, entropy, density, and volume, and second-order quantities such as heat capacity, isothermal compressibility, and isobaric thermal expansion coefficient. These response functions exhibit discontinuities or divergences at the phase transition. If encountering issues of insufficient experimental information, one must construct artificial target functions, such as step functions for energy and Gaussian peaks for heat capacity, introducing arbitrariness into the refinement that undermines objectivity.

Lee-Yang theory~\cite{yang1952statistical,lee1952statistical,fisher1965lectures} offers an elegant resolution. The theory establishes that phase transitions are encoded in the distribution of partition function zeros in the complex plane of the thermodynamic state variables (defined as Lee-Yang zeros), with these zeros approaching the real axis at the transition point in the thermodynamic limit. This suggests a fundamentally different refinement strategy: optimize the force field so that the partition function itself, rather than its derivatives, reflects the correct phase behavior. The recent development of practical schemes for computing Lee-Yang zeros from $NpT$ MD simulations~\cite{ouyang2024lee,liu2025determination} makes this strategy computationally feasible.

In this work, we propose a general framework for phase-diagram-guided force field refinement based on Lee-Yang theory. The central insight is that Lee-Yang theory provides a clear, general optimization target: at the phase transition, the corresponding Lee-Yang zero causes the partition function to vanish, so the partition function modulus itself, its zeroth-order information, can serve as the characterizing signal, avoiding the numerical difficulty of explicitly solving for the zeros. This yields a closed optimization loop: the partition function modulus evaluated from sampling trajectories forms a differentiable loss function, and gradient-based optimization refines force field parameters via automatic differentiation. Critically, this approach targets the partition function itself rather than system-specific surrogate properties, making it universal across different types of discontinuous phase transitions and material systems.

We now develop the theoretical foundation of our approach. In the $NpT$ ensemble, the partition function is an integral over enthalpy $H = U + pV$:

\begin{gather}
Z(\beta, p) = \int \rho_p(H)\, e^{-\beta H}\,\mathrm{d}H,
\end{gather}

where $\beta = 1/k_\mathrm{B}T$ and $\rho_p(H)$ is the density of states at pressure $p$. The potential $U$ depends on force field parameters $\boldsymbol{\theta}$, embedding the parameter dependence into $\rho_p(H)$. \emph{Different $\boldsymbol{\theta}$ produce different enthalpy distributions, hence different partition functions.}

To compute $Z$ from MD trajectories, we follow the discretization scheme of Refs.~\cite{ouyang2024lee,liu2025determination}. In the target enthalpy interval $[H_\mathrm{min}, H_\mathrm{max}]$, define $H_k = H_0 + k\Delta H$ for $k = 0, \ldots, N-1$, where $H_0 = H_\mathrm{min}, H_{N-1} = H_\mathrm{max}$. Then, extending $\beta \to \tilde{\beta} \in \mathbb{C}$ to analytically continue the partition function, we obtain:

\begin{gather}
\tilde{\mathcal{Z}}(\tilde{\beta}, p) = \Delta H\, e^{-\tilde{\beta} H_0} \sum_{k=0}^{N-1} \mathcal{\rho}_p(H_k) \left[e^{-\tilde{\beta} \Delta H}\right]^k.
\end{gather}

Using $\mathcal{G}_p(H) = \mathcal{\rho}_p(H) e^{-\beta H}$ to express the enthalpy probability distribution under the ensemble condition ($\beta, p$), and absorbing constants, we obtain a polynomial:

\begin{gather}
\tilde{\mathcal{Z}}(\tilde{\beta}, p) \propto \sum_{k=0}^{N-1} \mathcal{G}_p(H_k) \, \tilde{y}^k, \quad \tilde{y} = e^{-(\tilde{\beta} - \beta)\Delta H},
\end{gather}

where the polynomial coefficients $\mathcal{G}_p(H_k)$ are the enthalpy probability density at discrete enthalpy values $H_k$ obtained from MD simulations. \emph{This is the central connection: simulation data directly encodes the Lee-Yang zeros through the polynomial coefficients.}

The dependence chain from force field parameters to phase behavior is now explicit:
$$
\boldsymbol{\theta} \;\rightarrow\; \tilde{\mathcal{Z}}_{\boldsymbol{\theta}}(\tilde{\beta}, p) \;\rightarrow\; \text{Lee-Yang zeros} \;\rightarrow\; \text{phase diagram}.
$$

\begin{figure}[t]
    \begin{minipage}[t]{0.02\columnwidth}
        \vspace{0pt}
        (a)
    \end{minipage}
    \hfill
    \begin{minipage}[t]{0.95\columnwidth}
        \vspace{0pt}
        \includegraphics[width=\columnwidth]{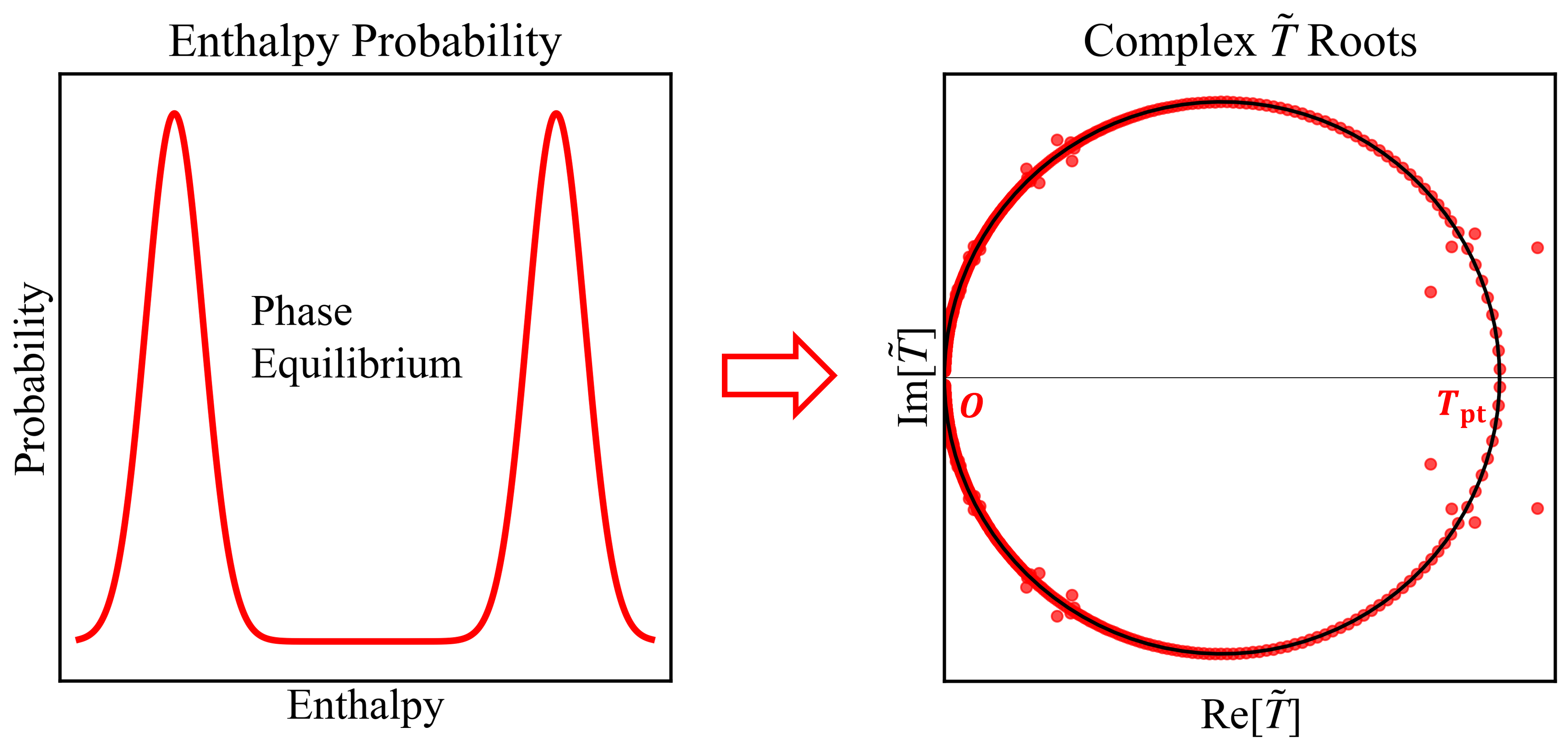}
    \end{minipage}
    
    \begin{minipage}[t]{0.02\columnwidth}
        \vspace{0pt}
        (b)
    \end{minipage}
    \hfill
    \begin{minipage}[t]{0.95\columnwidth}
        \vspace{0pt}
        \includegraphics[width=\columnwidth]{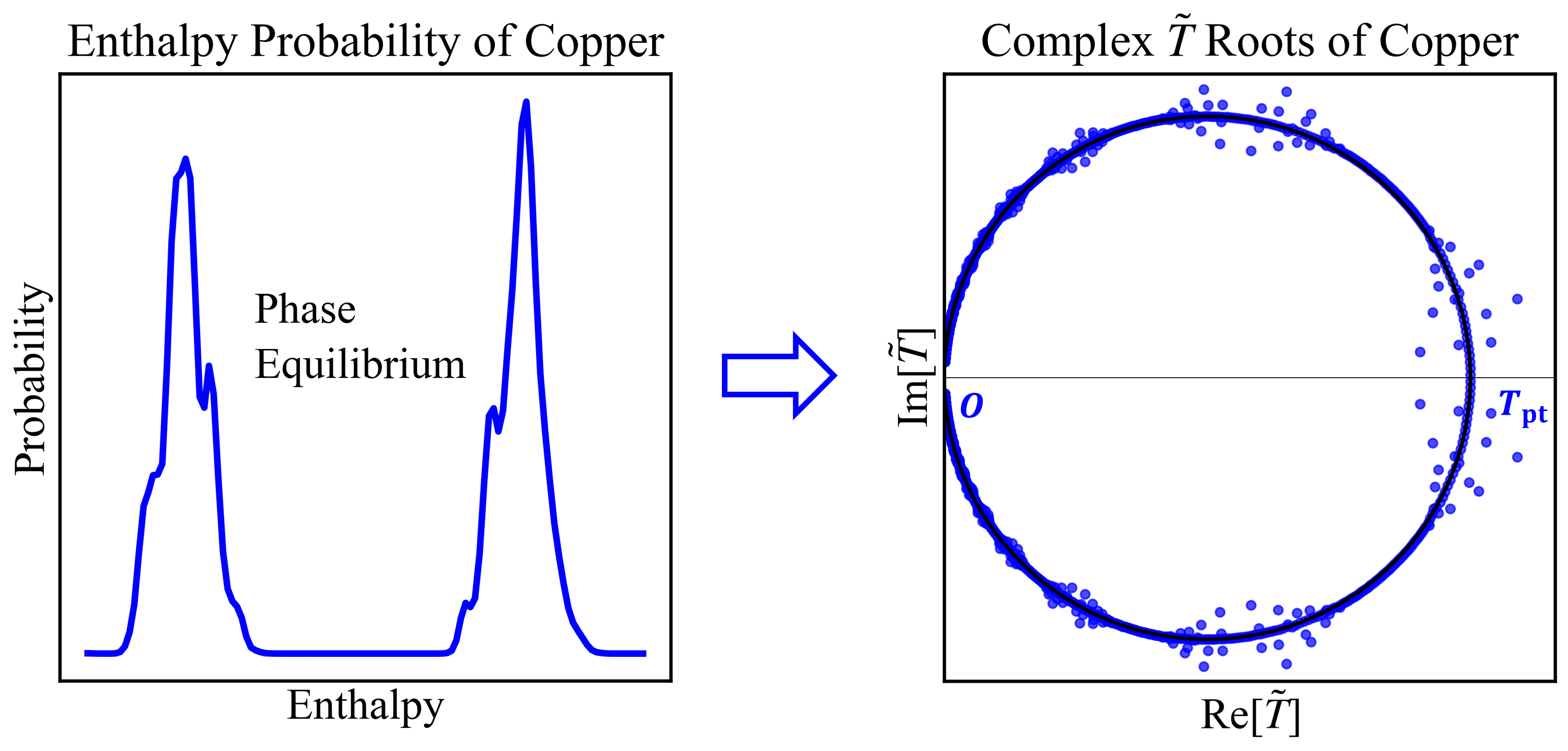}
    \end{minipage}
\caption{Schematic illustration of the Lee-Yang circle theorem in the complex temperature plane. (a) Ideal case: for a system exhibiting a double-peak equilibrium enthalpy distribution (left) with a phase transition at $T_\mathrm{pt}$, the Lee-Yang zeros most directly associated with the phase transition distribute on a circle with diameter connecting the origin and $T_\mathrm{pt}$~\cite{dong2026tracking} (right). (b) A realistic case for Cu, confirming this geometric picture.}
\label{fig:zero-circle}
\end{figure}

The Lee-Yang circle theorem provides the geometric foundation for inverting this chain. Originally proved for the grand canonical ensemble~\cite{yang1952statistical,lee1952statistical}, it has been extended to the $NVT$ temperature~\cite{rocha2014identifying,costa2017energy} and the $NpT$ temperature plane~\cite{liu2025determination}: for a system possessing an equilibrium enthalpy distribution like that depicted in the ideal case of Fig.~\ref{fig:zero-circle}(a) at pressure $p$ with a phase transition at $T_\mathrm{pt}$, the partition function zeros most directly associated with the phase transition lie on a circle in $\tilde{T}$-space with diameter $(0, T_\mathrm{pt})$~\cite{dong2026tracking} (named as Lee-Yang circle in the following), as shown in Fig.~\ref{fig:zero-circle}(a). In the thermodynamic limit, the Lee-Yang edge, defined as the zero with smallest imaginary part, approaches $T_\mathrm{pt}$ on the real axis. In Fig.~\ref{fig:zero-circle}(b) a realistic case for Cu confirms this geometric picture. \textbf{One target transition temperature corresponds to one circle of zeros.}

This geometry directly motivates a loss function. When the predicted zeros coincide with the target circle, the partition function modulus $|\tilde{\mathcal{Z}}|$ becomes small at points on this circle, vanishing in the thermodynamic limit. We therefore propose to refine force fields by minimizing $|\tilde{\mathcal{Z}}|$ at target points on the Lee-Yang circle. Unlike our prior gas-liquid work~\cite{jin2026automatic}, which optimized against density distribution similarity, this approach targets the partition function directly: a fundamental, universal signal applicable to common phase transitions.

Before constructing the loss function, the raw partition function requires a normalization step to ensure that changes in $|\tilde{\mathcal{Z}}|$ reflect genuine shifts in the Lee-Yang zeros, not artifacts from the change of the total number of microstates.

In practice, the partition function modulus computed with the raw probability density $\mathcal{G}_p(H)$ at target condition $(\beta, p)$ is influenced by the underlying number of microstates from two aspects, the total number and distribution of microstates. Both aspects change when the force field parameters change, while only the latter is related to the Lee-Yang zeros. If we minimize $|\tilde{\mathcal{Z}}|$ directly, the modulus could decrease simply because the total number of microstates changes, not because the Lee-Yang zeros are approaching the target.

To eliminate this artifact and isolate the signal arising from the Lee-Yang zeros, we normalize the probability distribution:
\begin{gather}
\mathcal{G}_k^p = \frac{\mathcal{G}_p(H_k)}{\sum_{k=0}^{N-1} \mathcal{G}_p(H_k)},
\end{gather}
ensuring $\sum_{k=0}^{N-1} \mathcal{G}_k^p = 1$ and maintaining the dependence on the interval's relative weight. This normalization is equivalent to computing the ensemble average $\langle e^{-(\tilde{\beta}-\beta)H}\rangle_{\beta, p}$. The final normalized partition function is
\begin{gather}
\bar{\mathcal{Z}}(\tilde{\beta}, p) = \sum_{k=0}^{N-1} \mathcal{G}_k^p \, \tilde{y}^k.
\end{gather}

\begin{figure}[t]
\includegraphics[width=\columnwidth]{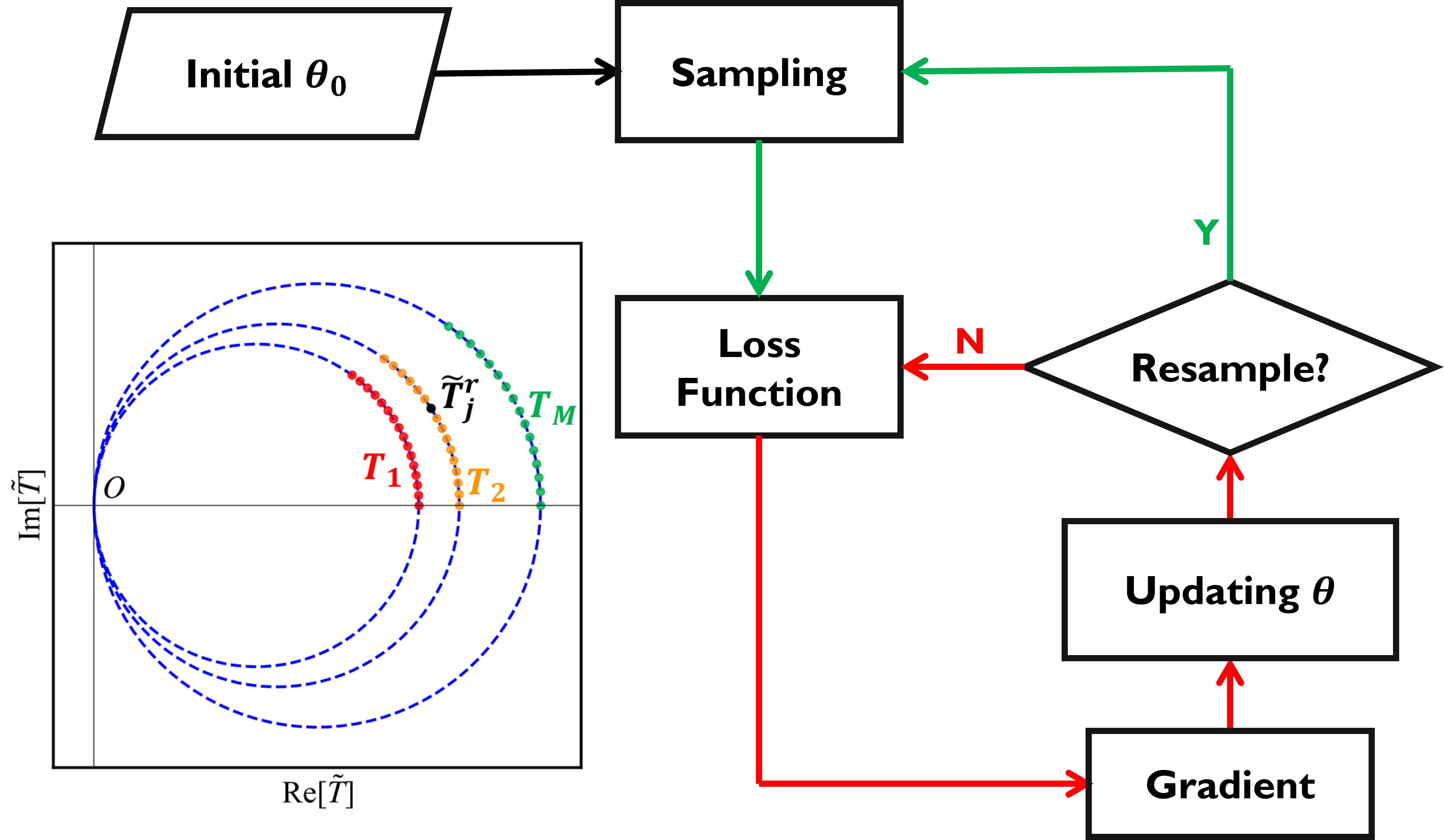}
\caption{Schematic workflow of the Lee-Yang-guided force field refinement framework. The left inset depicts points $\tilde{T}_j^r$ on the Lee-Yang circle, used to evaluate the partition function modulus and form a differentiable loss function minimized via automatic differentiation. Points of different colors correspond to different Lee-Yang zero circles for distinct target thermodynamic conditions.}
\label{fig:workflow}
\end{figure}

We now construct the loss function $\mathcal{L}(\boldsymbol{\theta})$ that drives force field refinement. For each target system at pressure $p_r$ with known phase transition temperature $T_r$ ($M$ target thermodynamic conditions are indexed by $r$, which could be from experiments), we select $P$ evaluation points in the complex temperature plane:
\begin{gather}
\tilde{T}_j^r = \frac{T_r}{2}\bigl[1 + \cos(j\Delta\theta) + i\sin(j\Delta\theta)\bigr], \quad j = 0, \ldots, P-1.
\end{gather}
These points lie on the upper arc of the Lee-Yang circle (zeros are symmetric about the real axis), starting from $T_r$ and moving into the complex plane with increasing $j$, as shown in the inset of Fig.~\ref{fig:workflow}, which are close to the real axis and more physically meaningful. At these $\tilde{T}_j^r$, the partition function modulus $|\bar{\mathcal{Z}}|$ approaches zero when the force-field-predicted Lee-Yang zeros coincide with the target circle.

The loss function aggregates over all target thermodynamic conditions and all evaluation points:
\begin{gather}
\mathcal{L}(\boldsymbol{\theta}) = \sum_{r=1}^{M} \sum_{j=1}^{P} \bigl|\bar{\mathcal{Z}}_{\boldsymbol{\theta}}(\tilde{\beta}_j^r, p_r)\bigr|^m,
\end{gather}
with exponent $m = 2$ chosen after preliminary numerical testing. Minimizing $\mathcal{L}(\boldsymbol{\theta})$ with respect to $\boldsymbol{\theta}$ drives the predicted Lee-Yang zeros toward the target circle, thereby matching the phase diagram. \emph{A key practical advantage is that no explicit zero-finding is required}; the partition function modulus itself serves as the signal, avoiding the numerical effort of polynomial root-solving.

Combined with the discretization and normalization established above, the optimization problem $\boldsymbol{\theta}^* = \arg\min_{\boldsymbol{\theta}} \mathcal{L}(\boldsymbol{\theta})$ constitutes a well-defined computational task.

We now describe the practical pipeline for acquiring the enthalpy distributions and implementing the differentiable optimization.

Solid-liquid phase transitions present a fundamental sampling challenge: the free energy barrier separating the two phases prevents conventional MD from exploring both basins on accessible timescales. To overcome this, we employ the On-the-fly Probability Enhanced Sampling (OPES) method~\cite{invernizzi2020opes}, which reconstructs the probability distribution $P(\mathbf{s})$ along selected collective variables via on-the-fly weighted kernel density estimation. The bias potential is constructed from the ratio of the unbiased and target distributions as $V(\mathbf{s}) = (1/\beta) \log [P(\mathbf{s}) / p^{\mathrm{tg}}(\mathbf{s})]$, with the well-tempered target $p^{\mathrm{tg}}(\mathbf{s}) \propto [P(\mathbf{s})]^{1/\gamma}$ used here. This bias flattens the underlying free energy landscape, enabling efficient barrier crossing while preserving rigorous reweightability to the unbiased ensemble.

For each target pressure $p_r$, we perform OPES $NpT$ MD simulations at several sampling temperatures bracketing the expected phase transition region. Configurations $\{\{\mathcal{X}_{l,i}\}_{i=1}^S\}_{l=1}^L$ from all $L$ sampling ensembles are assigned weights $W_{l,i}^r$ under the Multistate Bennett Acceptance Ratio (MBAR) framework~\cite{shirts2008mbar}, yielding the reweighted enthalpy probability distribution at any target temperature $T_r$:

\begin{gather}
\mathcal{G}_r(H) = \sum_{l=1}^{L} \sum_{i=1}^{S} W_{l,i}^r \, \delta\bigl(H_r(\mathcal{X}_{l,i}) - H\bigr).
\end{gather}

The target enthalpy interval and its discretization is detailed in the Supplemental Material Section S4.

The loss function $\mathcal{L}(\boldsymbol{\theta})$ is minimized with respect to the force field parameters $\boldsymbol{\theta}$ using the DMFF framework~\cite{wang2023dmff}, which leverages JAX~\cite{jax2024github} automatic differentiation for end-to-end gradient computation. The full differentiable chain is detailed in the Supplemental Material Section S2.

The refinement proceeds iteratively as shown in Fig.~\ref{fig:workflow}: (i)~setting the initial parameters $\boldsymbol{\theta}_0$, (ii)~sampling at the current parameters, (iii)~computing MBAR weights and constructing $\mathcal{G}_k^r$ at all target conditions, and evaluating $\mathcal{L}(\boldsymbol{\theta})$ and $\nabla_{\boldsymbol{\theta}} \mathcal{L}$ via DMFF, (iv)~updating $\boldsymbol{\theta}$ via gradient-based optimization method, (v)~resampling periodically to correct for MBAR reweighting error accumulation as $\boldsymbol{\theta}$ drifts from the reference used in the sampling.

This work considers two model systems. The Lennard-Jones (L-J) system covers both gas-liquid and solid-liquid phase transitions for force field refinement, while the Cu system focuses on solid-liquid transitions. Details of the sampling methods and convergence validation are provided in the Supplemental Material Section S1.

The L-J system consists of 256 particles, with gas-liquid target pressures $p = 0.06, 0.07, 0.08, 0.09, 0.10$~a.u. and solid-liquid target pressures $p = 0.20, 0.60, 1.00, 1.40$~a.u.. The Cu system employs an embedded-atom method (EAM) potential~\cite{daw1984embedded,mishin2001cu} with 500 atoms. More details of Cu EAM potential parameters are available in the Supplemental Material Section S3. For the simulated reference target, target pressures are $p = 1.00, 3.00, 5.00, 7.00$~GPa; for the experimental target, the pressures are $p = 1.73, 2.97, 4.69, 5.68$~GPa.

We validate the method in two stages. First, for both L-J and Cu, the refinement is tested against simulated reference phase diagrams obtained at well-converged force field parameters, serving as a controlled benchmark to verify that the Lee-Yang-guided optimization correctly recovers the known phase boundaries. Second, for Cu, the refinement is applied using experimental phase diagram data~\cite{errandonea2010cu} as the target, demonstrating the framework's practical applicability when an experimental target is provided. In each case the refinement starts from force field parameters perturbed from the target values and proceeds until the loss function converges. Full details of the optimization parameters are provided in the Supplemental Material Section S4.

The L-J system serves as a benchmark for $NpT$-ensemble $p$-$T$ phase diagrams, with refinement targeting both gas-liquid and solid-liquid phase transitions simultaneously.

\begin{figure}[t]
    \begin{minipage}[t]{0.02\columnwidth}
        \vspace{0pt}
        (a)
    \end{minipage}
    \hfill
    \begin{minipage}[t]{0.95\columnwidth}
        \vspace{0pt}
        \begin{overpic}[width=\columnwidth]{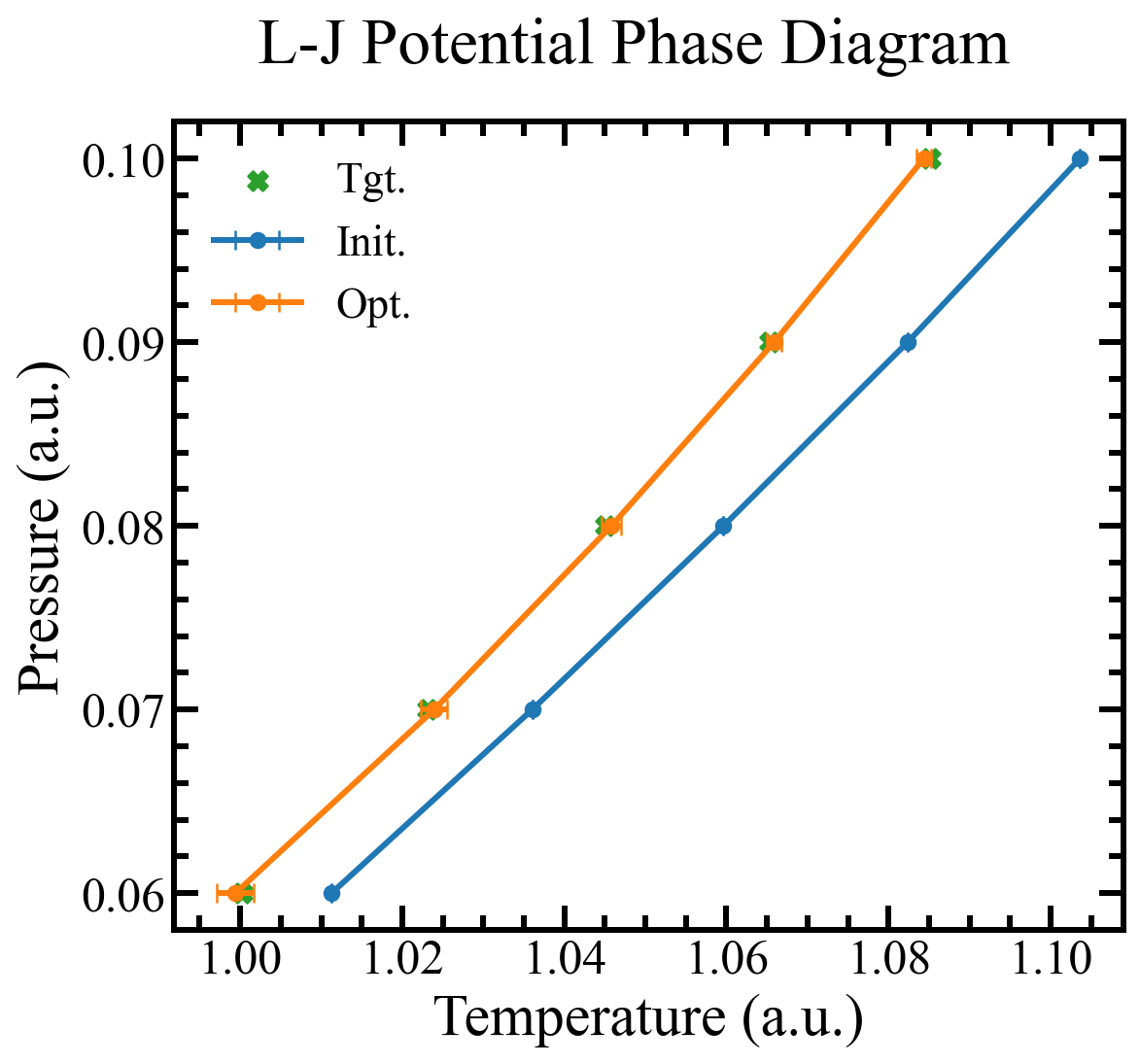}
            \put(55, 12){\includegraphics[width=0.5\columnwidth]{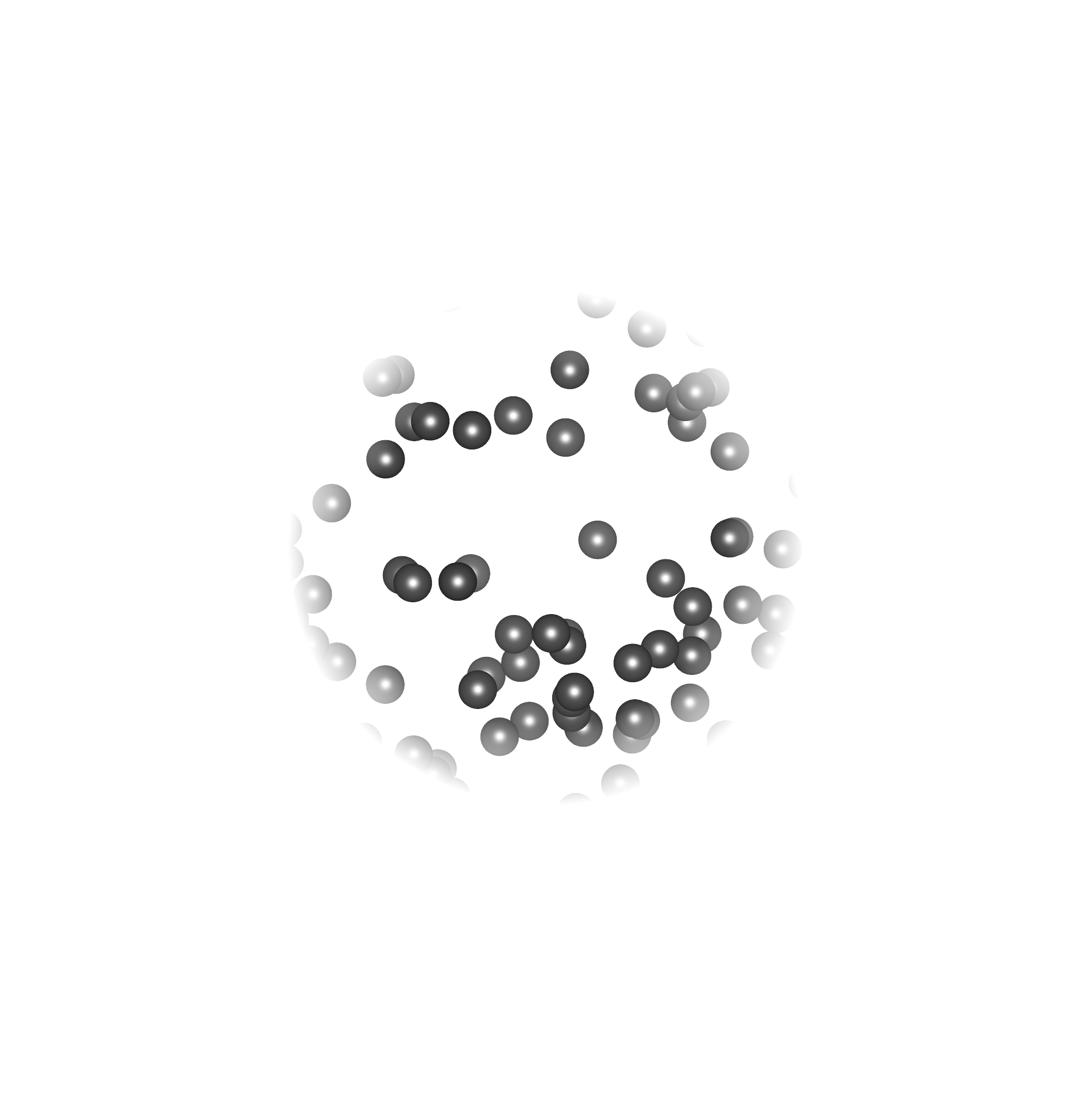}}
            \put(78, 20){(g)}
            \put(6, 22){\includegraphics[width=0.5\columnwidth]{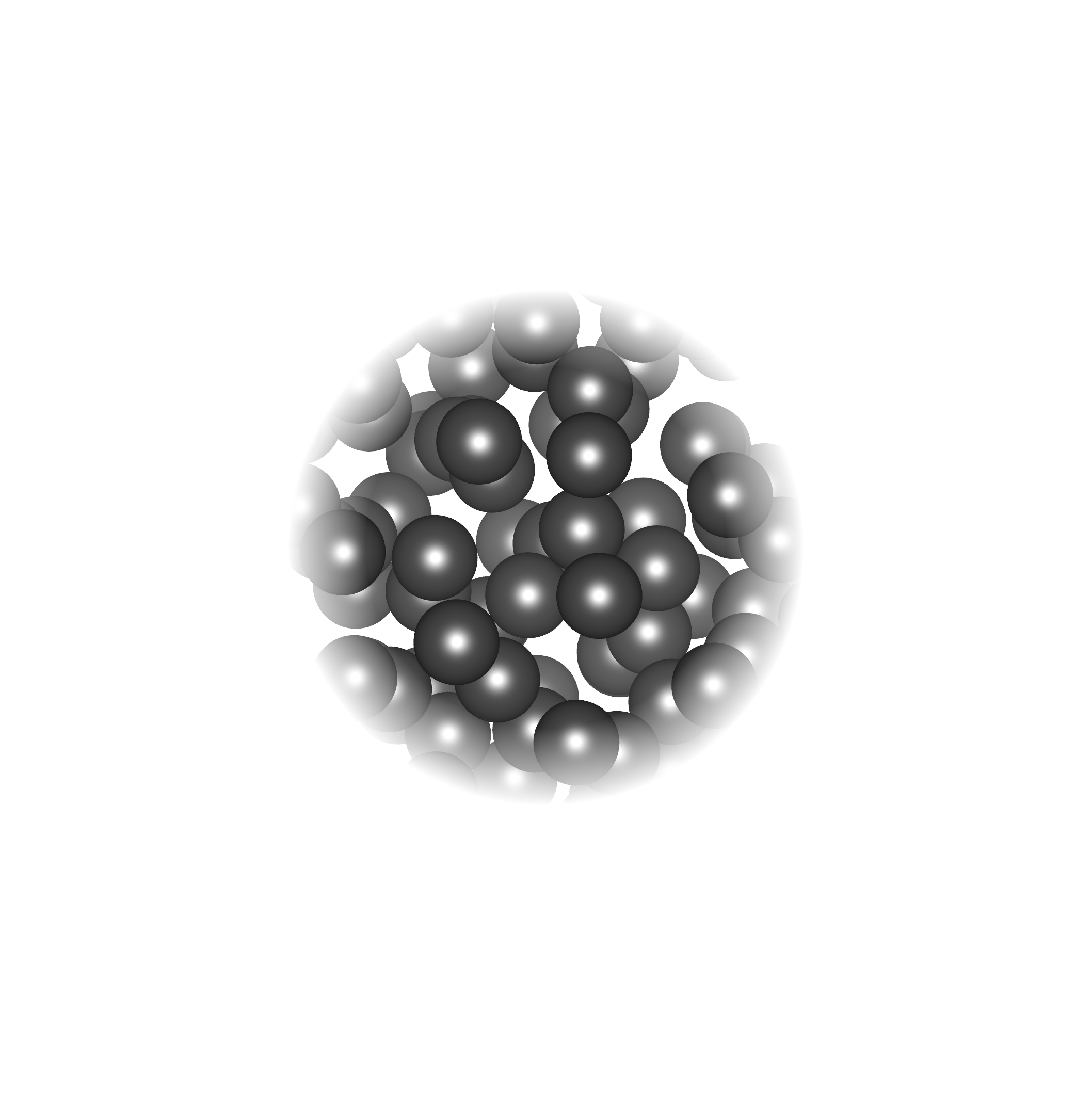}}
            \put(29, 30){(l)}
        \end{overpic}
    \end{minipage}
    
    \begin{minipage}[t]{0.02\columnwidth}
        \vspace{0pt}
        (b)
    \end{minipage}
    \hfill
    \begin{minipage}[t]{0.95\columnwidth}
        \vspace{0pt}
        \begin{overpic}[width=\columnwidth]{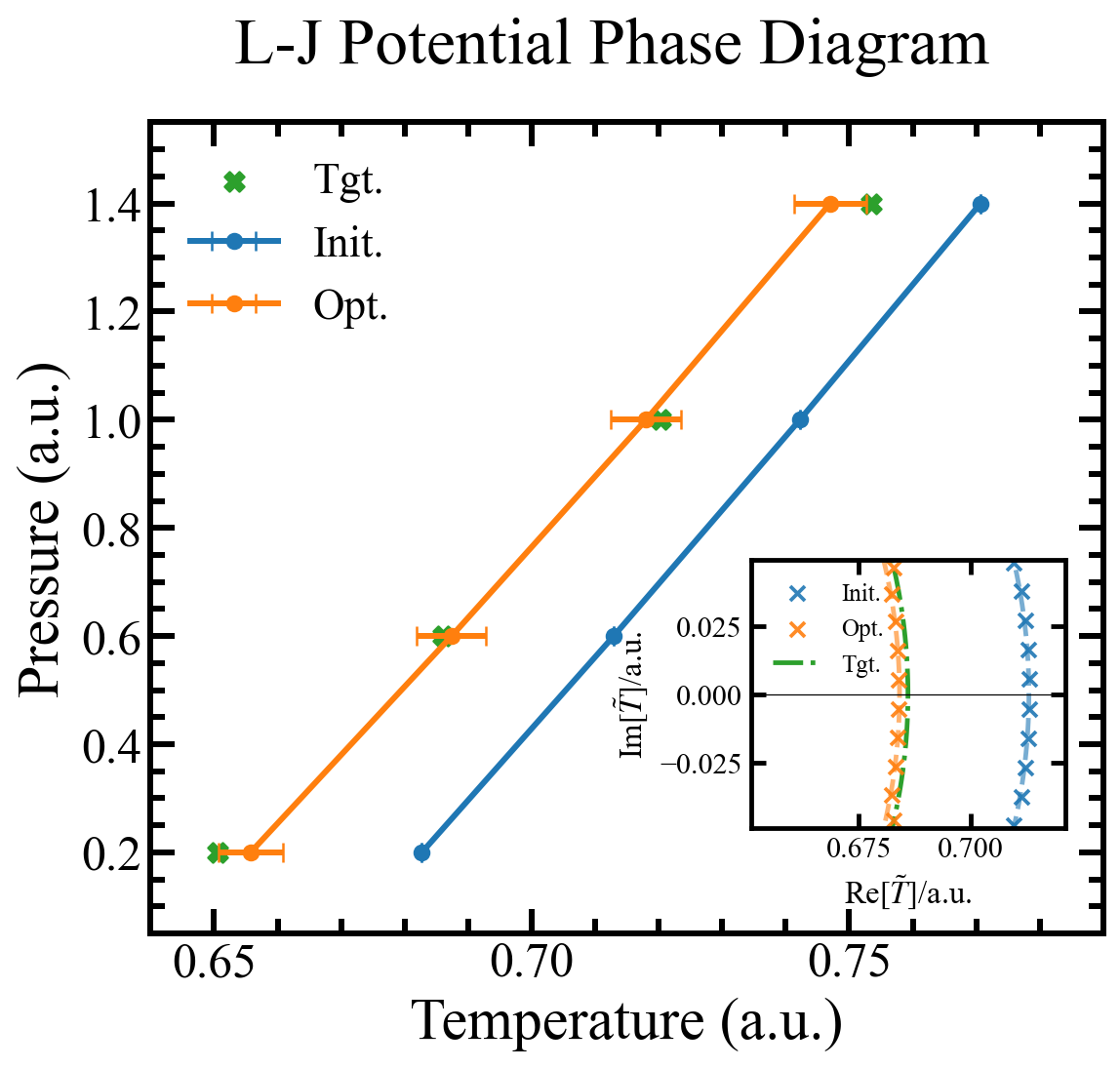}
            \put(60, 36){\includegraphics[width=0.5\columnwidth]{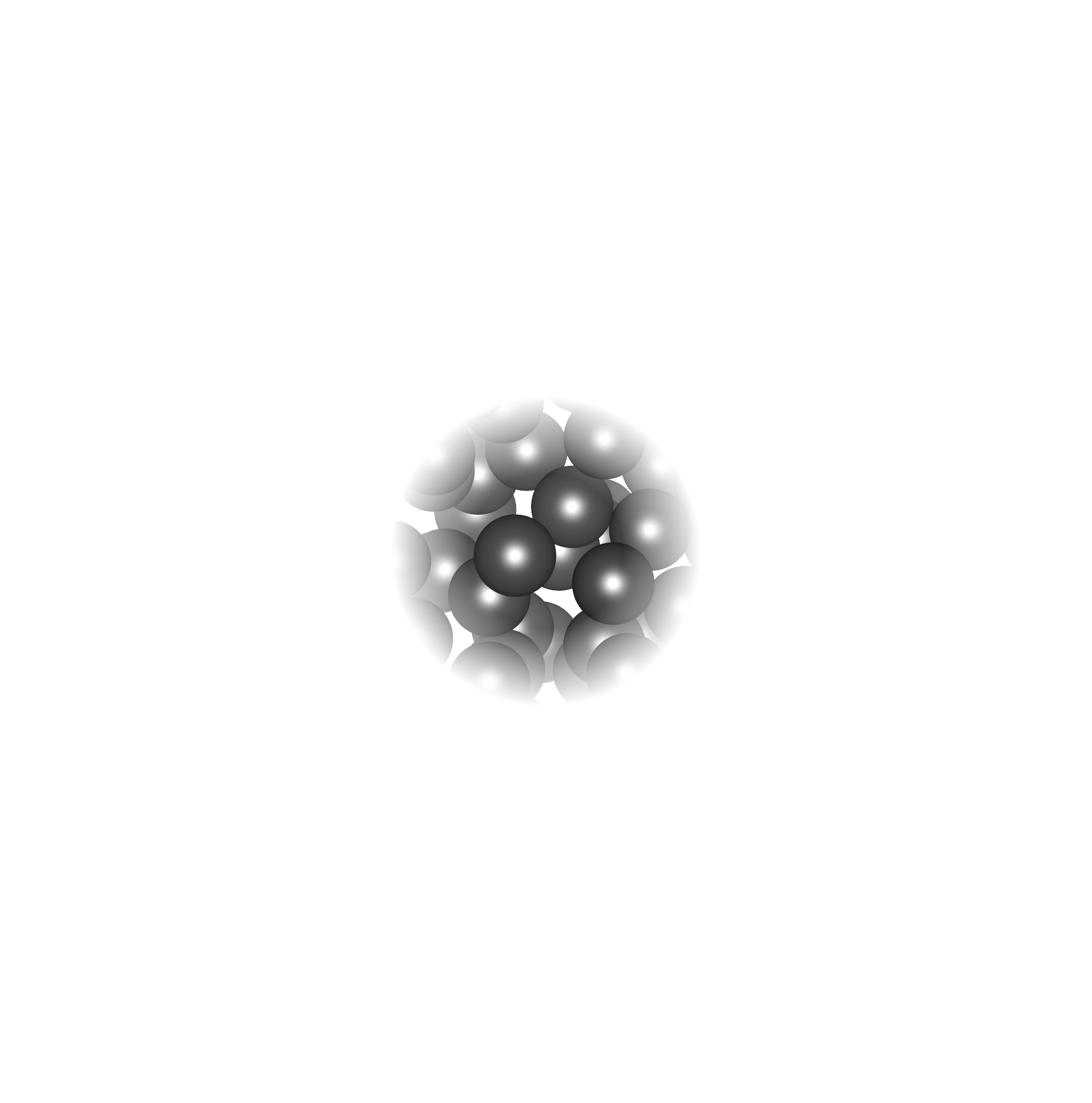}}
            \put(83, 50){(l)}
            \put(6, 28){\includegraphics[width=0.5\columnwidth]{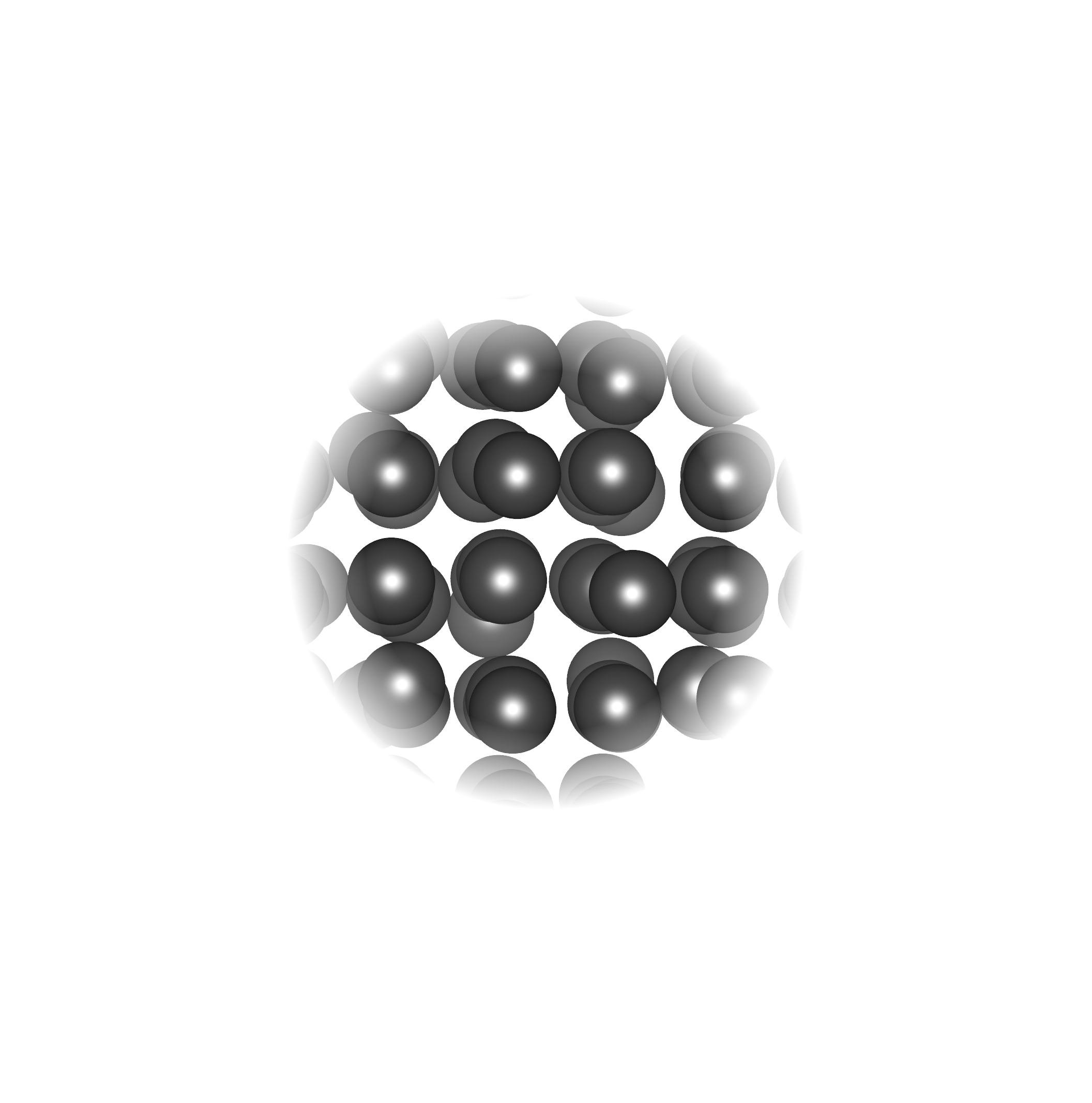}}
            \put(29, 36){(s)}
        \end{overpic}
    \end{minipage}
\caption{L-J system phase diagram optimization results. (a) Gas-liquid coexistence curves predicted by the initial force field (blue curve with points), refined force field (orange curve with points), and the target phase transition points (green crosses). (b) Solid-liquid coexistence curves with the same color scheme. The inset in (b) shows the evolution of Lee-Yang zeros near the real axis in the complex temperature plane at $p = 0.60$~a.u.: initial force field (blue crosses), refined force field (orange crosses), and the target Lee-Yang circle (green dashed line). Structures in figures are snapshots from simulations. (Same in Fig.~\ref{fig:cu-results}.)}
\label{fig:lj-results}
\end{figure}

Fig.~\ref{fig:lj-results}(a) presents the gas-liquid coexistence curves obtained with the initial and refined force fields, compared to the well-converged reference coexistence data. The refined force field reproduces the binodal with visibly reduced deviation across the full pressure range. Fig.~\ref{fig:lj-results}(b) presents the solid-liquid phase diagram, where the melting temperatures at the four target pressures are significantly closer to the reference values after refinement. The inset shows the evolution of Lee-Yang zeros near the real axis in the complex temperature plane at $p = 0.60$~a.u. (More results are shown in Supplemental Material Section S5). Before optimization, the zeros lie away from the target Lee-Yang circle. As the loss function $\mathcal{L}(\boldsymbol{\theta})$ decreases, the zeros progressively converge toward the Lee-Yang circle, and the real part of the Lee-Yang edge approaches the target transition temperature. Notably, a single set of refined L-J parameters simultaneously improves both phase transitions, demonstrating that the Lee-Yang-guided loss function captures the thermodynamic consistency requirements across different types of phase boundaries. The quantitative results are presented in detail in the Supplemental Material Section S7. For gas-liquid coexistence, the mean absolute error (MAE), defined as the average absolute deviation from the target phase transition temperature across all pressures, decreases from 0.0148 to 0.0007~a.u. (95.1~\% improvement). For solid-liquid transitions, the MAE decreases from 0.0246 to 0.0038~a.u. (84.4~\% improvement). These results confirm the efficacy and efficiency of the Lee-Yang-guided refinement framework for different types of phase transitions.

We next demonstrate the transferability of the method to metallic bonding using an EAM potential for Cu. We perform two sets of refinements.

\begin{figure}[t]
    \begin{minipage}[t]{0.02\columnwidth}
        \vspace{0pt}
        (a)
    \end{minipage}
    \hfill
    \begin{minipage}[t]{0.95\columnwidth}
        \vspace{0pt}
        \begin{overpic}[width=\columnwidth]{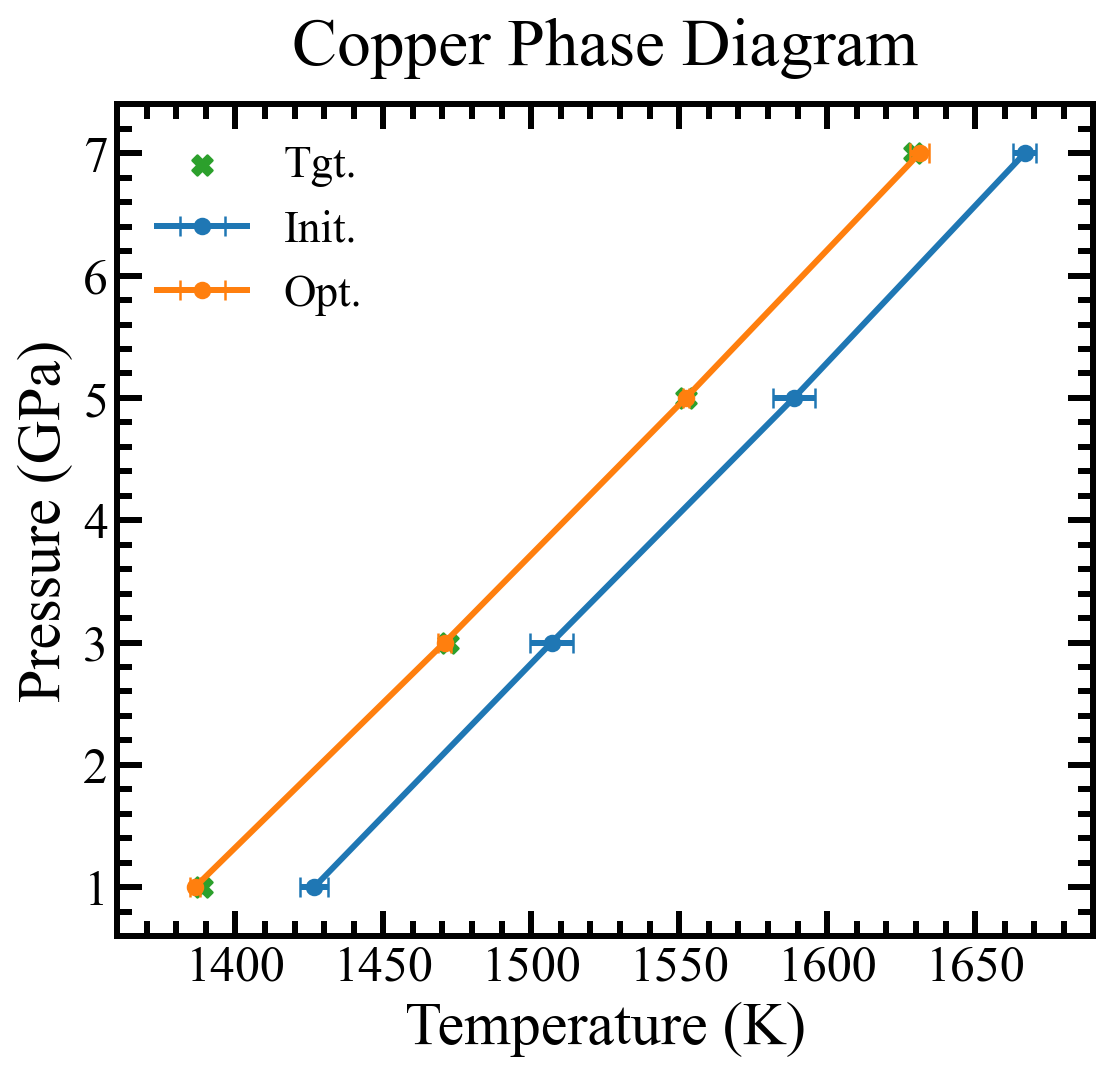}
            \put(53, 12){\includegraphics[width=0.5\columnwidth]{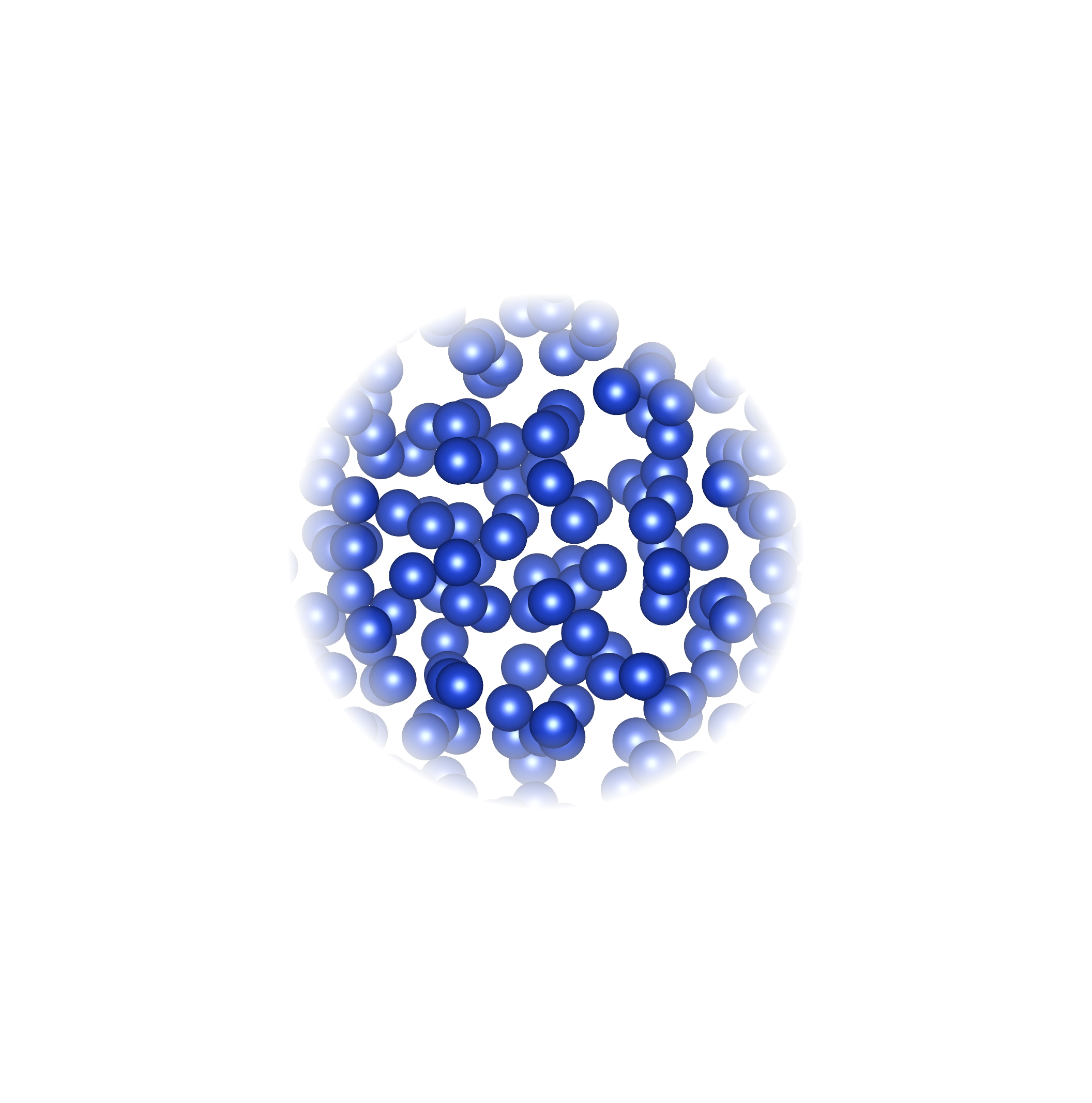}}
            \put(76, 20){(l)}
            \put(5, 29){\includegraphics[width=0.5\columnwidth]{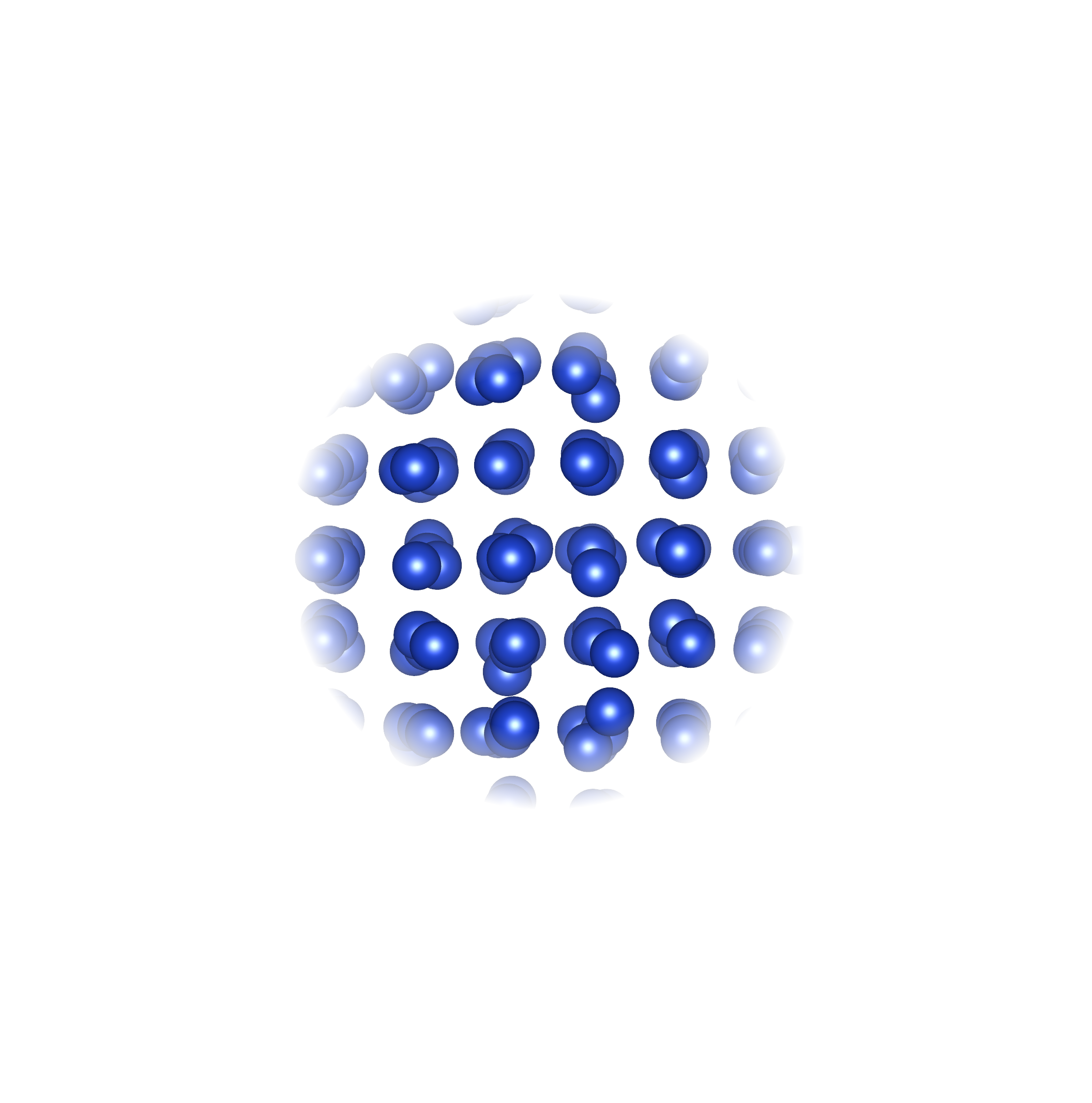}}
            \put(28, 37){(s)}
        \end{overpic}
    \end{minipage}
    
    \begin{minipage}[t]{0.02\columnwidth}
        \vspace{0pt}
        (b)
    \end{minipage}
    \hfill
    \begin{minipage}[t]{0.95\columnwidth}
        \vspace{0pt}
        \includegraphics[width=\columnwidth]{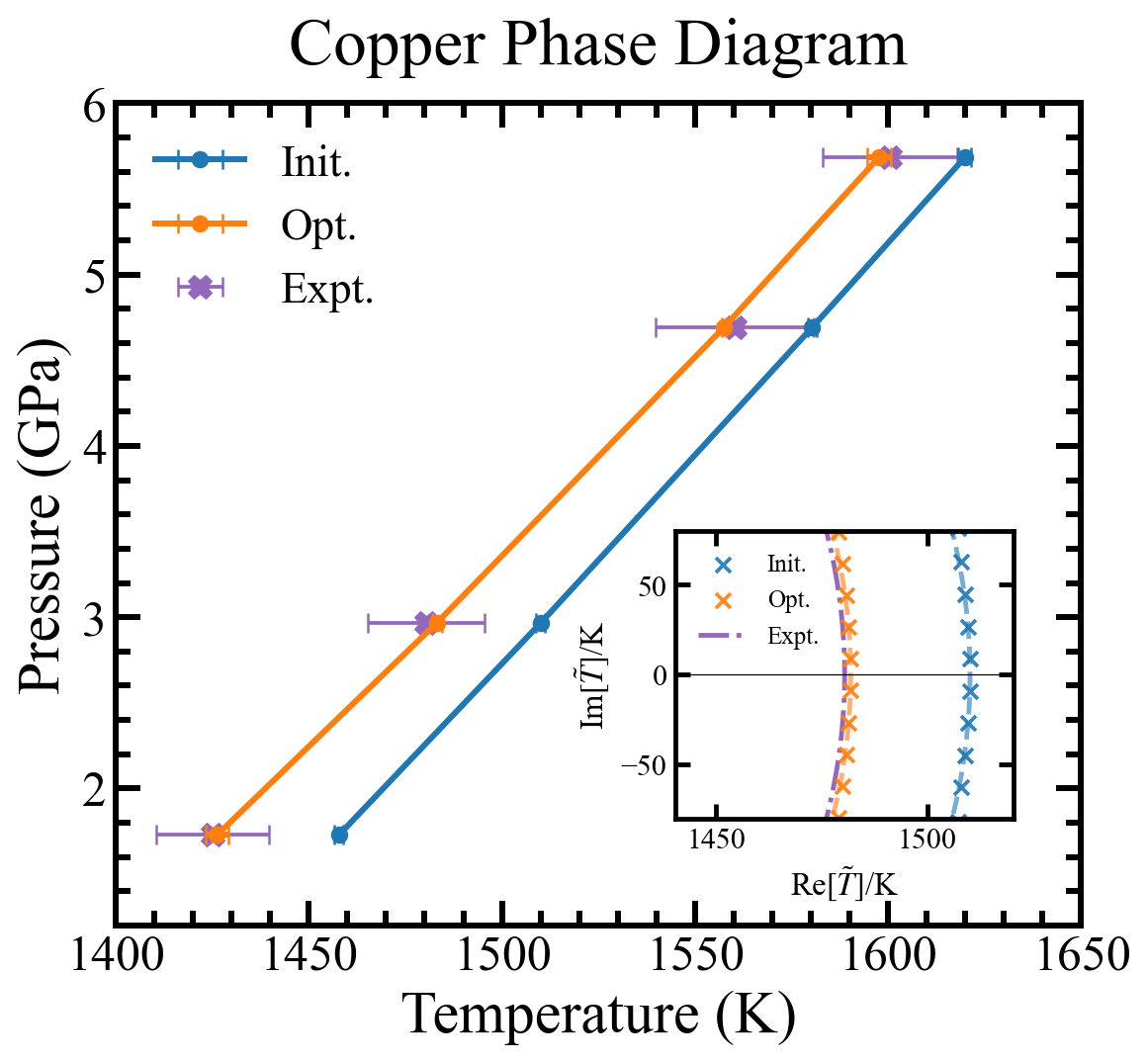}
    \end{minipage}
\caption{Cu EAM system optimization results. (a) Melting curve for the simulated reference target, comparing the initial force field (blue curve with points), refined force field (orange curve with points), and reference data (green crosses). (b) Melting curve for the experimental target, comparing the initial force field (blue curve with points), refined force field (orange curve with points), and experimental data (violet crosses)~\cite{errandonea2010cu}. The inset shows the evolution of Lee-Yang zeros near the real axis in the complex temperature plane at pressure $p = 2.97$~GPa for the experimental target: initial force field (blue crosses), refined force field (orange crosses), and the target Lee-Yang circle (violet dashed line).}
\label{fig:cu-results}
\end{figure}

Fig.~\ref{fig:cu-results}(a) compares the initial and refined melting curves against the simulated reference target, and Fig.~\ref{fig:cu-results}(b) compares them against the experimental target~\cite{errandonea2010cu}. In both cases, the refined force field reproduces the target melting temperatures with substantially improved accuracy. The insets show the corresponding evolution of Lee-Yang zeros near the real axis at a representative target condition (More results are shown in Supplemental Material Section S5). The zeros move from positions distant from the target Lee-Yang circle toward tight clustering on it as refinement proceeds, confirming that the Lee-Yang-guided loss function is effective for this metallic system as well. The quantitative results are presented in the Supplemental Material Section S7. For the simulated reference target, the MAE decreases from 36.7~K to 1.4~K (96.1~\% improvement). For the experimental target, the MAE decreases from 25.4~K to 2.4~K (90.6~\% improvement).

For the refinement against experimental data, we further validate the refined force field against thermodynamic observables not used as optimization targets.

\begin{figure}[t]
    \begin{minipage}[t]{0.02\columnwidth}
        \vspace{0pt}
        (a)
    \end{minipage}
    \hfill
    \begin{minipage}[t]{0.95\columnwidth}
        \vspace{0pt}
        \includegraphics[width=\columnwidth]{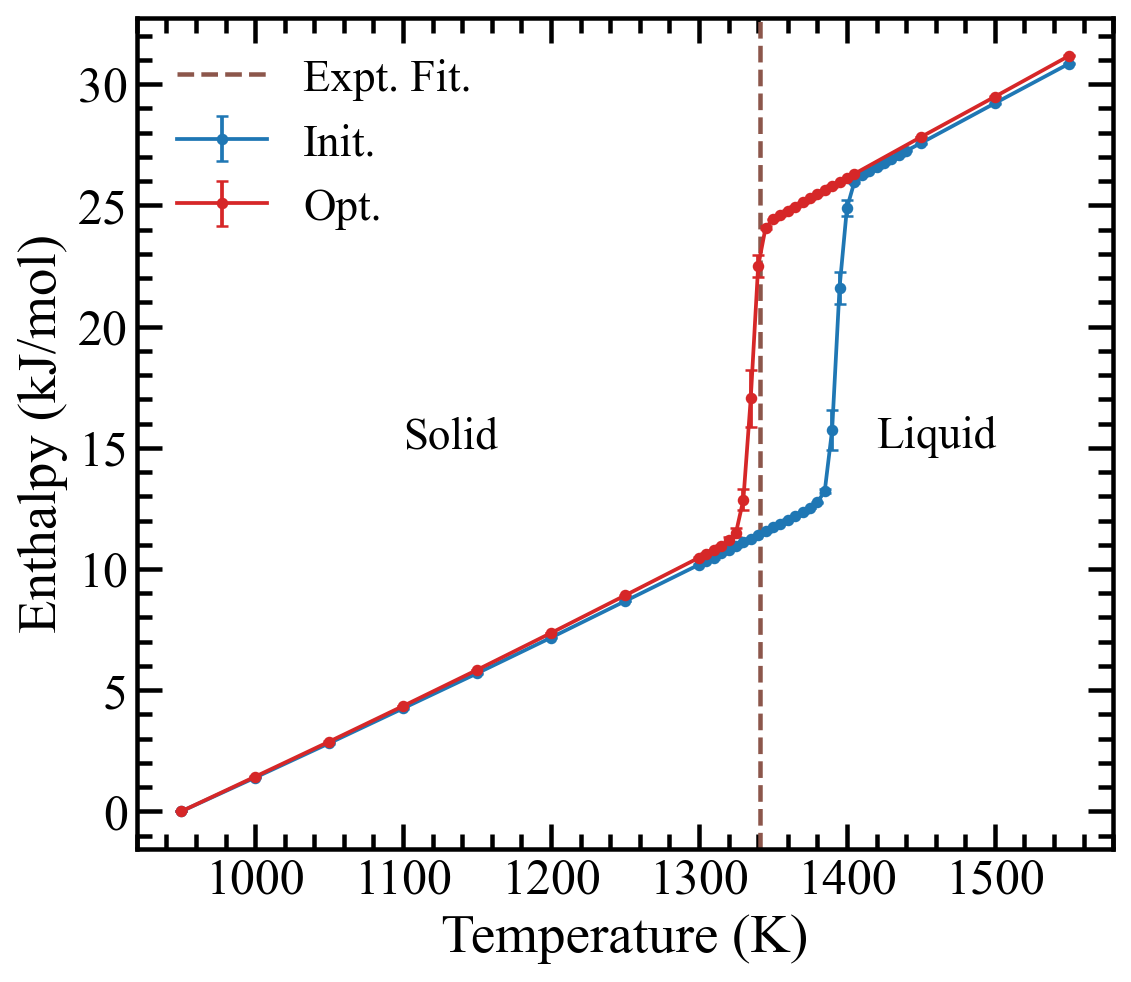}
    \end{minipage}
    
    \begin{minipage}[t]{0.02\columnwidth}
        \vspace{0pt}
        (b)
    \end{minipage}
    \hfill
    \begin{minipage}[t]{0.95\columnwidth}
        \vspace{0pt}
        \includegraphics[width=\columnwidth]{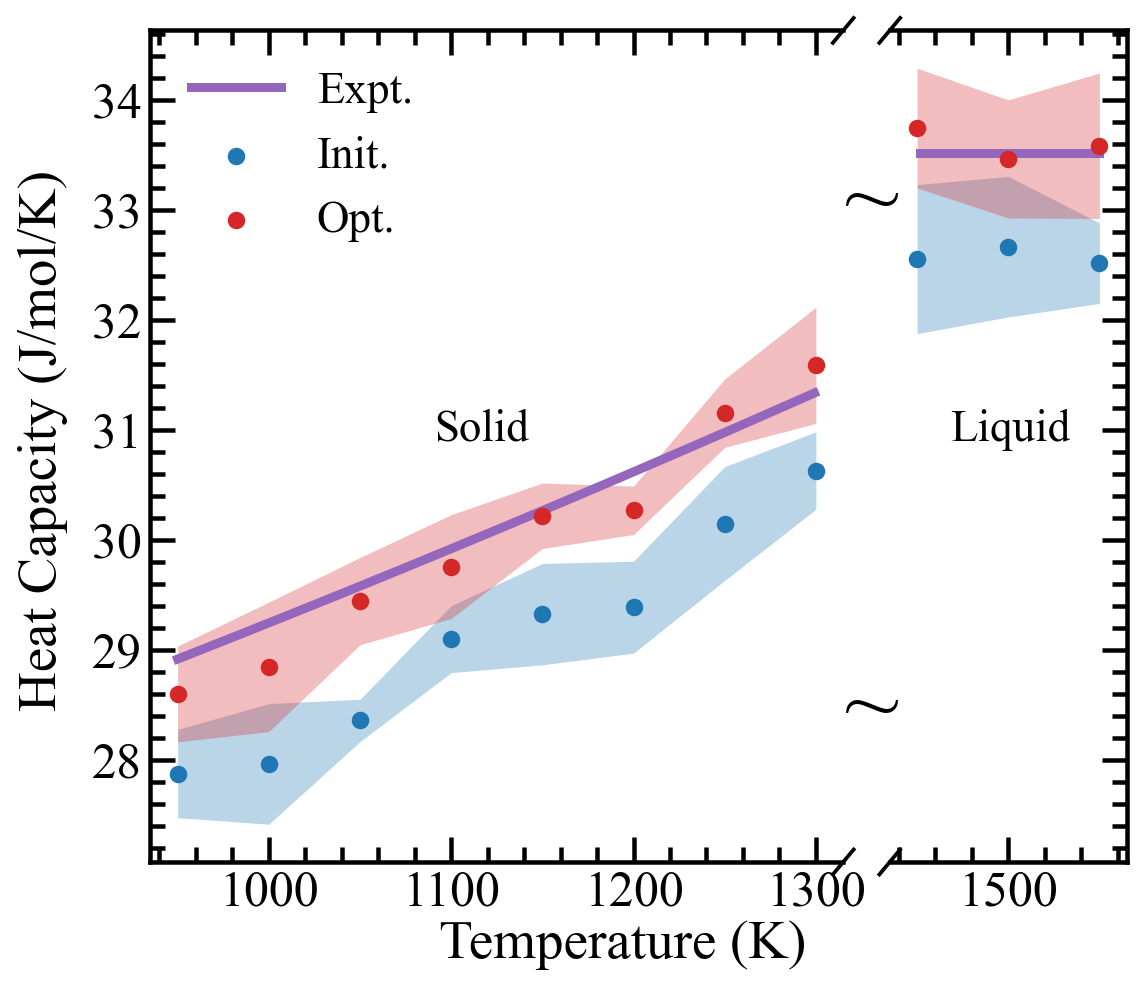}
    \end{minipage}
\caption{Thermodynamic validation for Cu at $p = 1$~bar. (a) Enthalpy $H(T)$ and (b) isobaric heat capacity $C_p(T)$ spanning the solid and liquid phases. In panel (a), the blue and red curves with points represent the enthalpies predicted by the initial and refined force fields, respectively. The brown dashed line denotes the ambient-pressure melting point obtained by extrapolating the experimental optimization targets~\cite{errandonea2010cu} via the Simon equation. In panel (b), the blue and red points represent the isobaric heat capacities predicted by the initial and refined force fields, respectively, with semi-transparent bands indicating the uncertainty of each point. The violet line shows the experimental data~\cite{kozyrev2023cu}. The phase transition region is omitted in panel (b) because experiment cannot adequately resolve the singular behavior of $C_p(T)$ near the transition.}
\label{fig:cu-thermo}
\end{figure}

Fig.~\ref{fig:cu-thermo} compares the enthalpy $H(T)$ [Fig.~\ref{fig:cu-thermo}(a)] and isobaric heat capacity $C_p(T)$ [Fig.~\ref{fig:cu-thermo}(b)] at $p = 1$~bar over a temperature range spanning the solid and liquid phases, for the initial force field, the refined force field, and experimental targets. For the enthalpy $H(T)$, both the initial and refined force fields exhibit the same qualitative behavior: $H(T)$ increases gradually with temperature in each phase and undergoes an abrupt jump at the melting transition, with the overall shape of the curve remaining largely unchanged. The refined force field, however, yields a melting temperature that is significantly closer to the ambient-pressure melting point obtained by extrapolating the experimental optimization targets~\cite{errandonea2010cu} via the Simon equation~\cite{simon1930melting,salter1954simon} (see Supplemental Material Section S6). For the isobaric heat capacity $C_p(T)$, Fig.~\ref{fig:cu-thermo}(b) omits the transition region and compares the solid and liquid phases separately, since experiment cannot adequately resolve the singular behavior of $C_p$ near the phase transition. In both phases, the refined force field yields $C_p$ values that are clearly closer to the experimental data~\cite{kozyrev2023cu} than those of the initial force field.

These results demonstrate that the Lee-Yang-guided framework exhibits promising transferability among different types of material systems, and that optimizing against phase diagram accuracy yields force fields with improved predictive power for related thermodynamic observables.

Several aspects of the proposed framework merit further discussion. First, evaluating the partition function at points on the Lee-Yang circle near the real axis is physically meaningful: these are the locations where the zeros are most directly associated with the phase transition. The circle geometry provides a natural, principled set of evaluation points parameterized by the target transition temperature alone, connecting the loss function directly to the phase behavior without introducing extraneous parameters.

Second, using the partition function modulus $|\bar{\mathcal{Z}}|$ rather than explicitly solving for individual Lee-Yang zeros offers two practical advantages. It avoids the numerical challenge of polynomial root-finding, which can be ill-conditioned for the enthalpy distributions obtained from finite MD sampling; moreover, the root-finding process does not lend itself naturally to a differentiable formulation of the loss function. In addition, a modulus-based loss integrates information from all zeros relevant to the phase transition, whereas explicit zero-finding may identify roots unrelated to the transition of interest. The modulus naturally weights the regions on the Lee-Yang circle where the physical zeros concentrate, providing a robust signal without the discontinuities or branch-switching issues inherent in tracking individual zeros across optimization steps.

At a broader level, the framework's foundation in the zeroth-order information of the partition function gives it a universality absent from existing force field refinement methods. It requires only the phase diagram as input; no order parameters, physical quantities, or system-specific choices are needed. This makes it applicable to common phase coexistence problems for which a target $p$-$T$ phase boundary can be specified, regardless of the nature of the competing phases or the complexity of the interatomic interactions.

However, we acknowledge that the current framework still has several limitations and extension directions for future work. The system size used in the optimization should be sufficiently large: for small simulation cells, the density of Lee-Yang zeros near the real axis may be too sparse, weakening the modulus signal that drives the refinement. As the thermodynamic limit is approached, the zero density near the real axis increases, strengthening the connection between the loss function and the phase transition. Additionally, several parameter choices introduce a degree of arbitrariness: the selection method and angular range of used points on the Lee-Yang circle, the number of enthalpy bins $N$, and the exponent $m$ or a potential polynomial combination of different powers in the loss function. While our numerical tests indicate that results are robust to reasonable variations, a systematic study of their influence would strengthen the framework. The present work uses a fixed resampling interval during optimization; for more complex systems where the MBAR reweighting error accumulates at a system-dependent rate, an adaptive resampling strategy based on effective sample size could improve efficiency and robustness. Extension to machine-learning (ML) force fields is a natural next step. The L-J potential has only 2 parameters and Cu EAM over 10, and the automatic differentiation backbone of the framework makes it reasonable to speculate that the approach can scale to the much larger parameter spaces of ML force fields, where direct phase-diagram-guided optimization would be particularly valuable given the difficulty of training ML potentials that accurately capture phase boundaries.

In conclusion, we have proposed a general framework for force field refinement based on Lee-Yang phase transition theory. The key insight is that the partition function modulus can serve as a direct, phase-diagram-guided optimization target, free from the need for system-specific order parameters or response properties. By connecting force field parameters to the partition function and constructing a differentiable loss function minimized via automatic differentiation, the framework provides a closed optimization loop that targets the phase diagram.

Validated on both an L-J potential covering gas-liquid and solid-liquid phase transitions and a Cu EAM potential targeting both the simulated and experimental melting curves, the phase diagrams reproduced by the refined force fields are significantly improved. For Cu, the refinement against experimental data simultaneously improves the predictions of enthalpy and heat capacity---thermodynamic observables outside the optimization target---demonstrating that phase-diagram-guided optimization yields force fields with broad predictive power. The framework is universal, applying to common phase coexistence, from model potentials to realistic materials systems.

This work demonstrates that Lee-Yang theory can serve as a practical tool for modern force field development. Extension to multi-component systems and machine-learning force fields, where the vast parameter space makes direct phase-diagram-guided optimization particularly valuable, is a natural next step.

\begin{acknowledgments}
The authors acknowledge funding support from the National Natural Science Foundation of China (Grant No.~92470114, No.~52273223), Ministry of Science and Technology of the People's Republic of China (Grant No.~2021YFB3800303), and DP Technology Corporation (Grant No.~2021110016001141). The computing resource of this work was provided by the Bohrium Cloud Platform, which was supported by DP Technology.
\end{acknowledgments}

\bibliographystyle{apsrev4-2}
\bibliography{references}
\end{document}